\documentclass[%
 aip,
 amsmath,amssymb,
 reprint,%
]{revtex4-1}

\usepackage{graphicx}
\usepackage{dcolumn}
\usepackage{bm}

\usepackage[utf8]{inputenc}
\usepackage[T1]{fontenc}
\usepackage{mathptmx}
\usepackage{etoolbox}

\makeatletter
\def\@email#1#2{%
 \endgroup
 \patchcmd{\titleblock@produce}
  {\frontmatter@RRAPformat}
  {\frontmatter@RRAPformat{\produce@RRAP{*#1\href{mailto:#2}{#2}}}\frontmatter@RRAPformat}
  {}{}
}%
\makeatother
\begin{document}

\preprint{AIP/123-APL}

\title[High-Q guided-mode resonance of a crossed grating with near-flat dispersion]{High-Q guided-mode resonance of a crossed grating with near-flat dispersion}
\author{Reuben Amedalor}
\affiliation{ 
Photonics Laboratory, Tampere University, FI-33720 Tampere, Finland
}%
\affiliation{Center for Photonics Sciences, University of Eastern Finland, FI-80101 Joensuu, Finland}
\author{Petri Karvinen}%
\affiliation{Center for Photonics Sciences, University of Eastern Finland, FI-80101 Joensuu, Finland
}%
\author{Henri Pesonen}
\affiliation{%
Center for Photonics Sciences, University of Eastern Finland, FI-80101 Joensuu, Finland
}%
\author{Jari Turunen}
\affiliation{%
Center for Photonics Sciences, University of Eastern Finland, FI-80101 Joensuu, Finland
}%
\author{Tapio Niemi}
\affiliation{ 
Photonics Laboratory, Tampere University, FI-33720 Tampere, Finland
}%
\author{Subhajit Bej}
\affiliation{ 
Photonics Laboratory, Tampere University, FI-33720 Tampere, Finland
}%
\email{subhajit.bej@tuni.fi}

\date{\today}

\begin{abstract}
Guided-mode resonances in diffraction gratings are manifested as peaks (dips) in reflection (transmission) spectra. Smaller resonance line widths (higher Q-factors) ensure stronger light-matter interactions and are beneficial for field-dependent physical processes. However, strong angular and spectral dispersion are inherent to such high-Q resonances. We demonstrate that a class of high-Q resonant modes (Q-factor >1000) exhibiting extraordinarily weak dispersion can be excited in crossed gratings simultaneously with the modes with well-known nearly linear dispersion. Furthermore, we show that the polarization of the incoming light can be adjusted to engineer the dispersion of these modes, and strong to near-flat dispersion or vice-versa can be achieved by switching between two mutually orthogonal linear polarization states. We introduce a semi-analytical model to explain the underlying physics behind these observations and perform full-wave numerical simulations and experiments to support our theoretical conjecture. The results presented here will benefit all applications that rely on resonances in free-space-coupled geometries.
\end{abstract}

\maketitle
\noindent Resonance anomalies of diffraction gratings attracted wide attention for over half a century \cite{petit1980}. At first, metallic gratings were shown to have sharp reflection dips associated with the excitation of surface modes, now known as surface plasmon polaritons \cite{Wood1902}. A different class of anomalies associated with guided-wave excitation in dielectric gratings has also been well studied \cite{Quaranta2018}. The gratings that exhibit this type of anomaly manifested as sharp peaks in reflection, and corresponding dips in transmission spectra, are known as resonant waveguide gratings (RWGs) or guided-mode resonance filters (GMRFs). The character and the spectral shape of the anomaly depend critically on the structural geometry, as well as on the arrival angle and the polarization state of the incident electromagnetic wave.

One of the most common geometries of a GMRF includes a linear binary grating on top of a thin-film optical waveguide, and resonance anomalies can be connected with the excitation of waveguide modes \cite{Wang93}. Diffraction from the grating enables phase matching of incident light with the propagating waveguide modes, which eventually radiate into free space and interfere with light in direct reﬂection and transmission (in $0^{th}$ diffraction order). Consequently, sharp peaks in reflection and dips in transmission are observed. GMRFs have been exploited for numerous applications such as filtering and beam splitting \cite{Liu98, Yoshiaki2008, Streshinsky2013, XU2017}, refractive index sensing \cite{Kikuta2001, Magnusson2011}, optical signal processing \cite{Doskolovich2014}, diffractive identiﬁcation \cite{Wu2007}, and wavelength division multiplexing \cite{Kintaka2010, Magnusson2012}.

GMRFs, with one-dimensional (1-D) periodicity, have been studied the most. The nature of diffraction from 1-D GMRFs depends on their illumination conditions. When the grating lines are perpendicular to the plane of incidence (POI), i.e., the plane containing the propagation vector of the incident harmonic wave and the unit vector normal to the grating surface, the propagation vectors of the diffracted waves lie in the POI. This excitation geometry is known as classical diffraction mounting. Whereas for conical diffraction mounting \cite{Moharam1995}, the grating lines form an arbitrary angle with the POI, and the diffracted wave vectors reside on the surface of a cone. The grating lines can also be parallel to the POI, and the situation is known as full-conical diffraction mounting. There are plenty of published works on the classical \cite{Rosenblatt1997, Saleem2012}, and conical  diffraction mounting geometries of 1-D GMRFs \cite{Lacour2003,Niederer:05}.

One crucial characteristic of a GMRF is its resonance linewidth which is often defined in terms of the Q-factor, defined as the ratio of the resonance peak wavelength to the resonance linewidth and usually scales inversely with the angular bandwidth of the incident light beam. Lemarchand \emph{et al.} proposed using a grating with two collinear periodicities to maintain high-Q and wider angular tolerance simultaneously \cite{Lemarchand1998}. Considerably larger angular acceptance could also be obtained under a full-conical diffraction mounting of a 1-D GMRF \cite{Ko:16}.

Two-dimensional (2-D) GMRFs, i.e., gratings with two noncollinear, in-plane lattice periodicities, are also extensively studied. One particular case of a 2-D grating is a crossed grating where the two grating lines are orthogonal to each other. Peng and Morris theoretically investigated the guided mode resonances in a crossed grating \cite{Peng:96:1}, and later experimentally demonstrated the resonant modes at normal incidence of light \cite{Peng:96:2}. Mizutani \emph{et al.} realized polarization independence using a 2-D grating with a rhombic lattice structure \cite{Mizutani:01}. Fehrembach \emph{et al.} developed a phenomenological theory for 2-D GMRFs \cite{Fehrembach:02}. Later, they introduced a perturbative approach to analyze the resonant modes at an oblique incidence \cite{Fehrembach:03}. Wang \emph{et al.} attempted to explain the origin of various spectral features of 2-D GMRFs with rectangular lattices \cite{Wang:19} and proposed design principles for polarization-independent 2-D GMRFs for non-normal incidence of a plane wave \cite{Wang2018}.

In this letter, we experimentally demonstrate that in a 2-D grating, a type of high-Q resonance with near-flat dispersion can be excited together with the typical dispersive resonant mode. Additionally, we show that the polarization state of light can control the dispersion characteristics of the resonances. Specifically, by switching between two mutually orthogonal linear polarization states, one can trigger a change from linear to near-zero dispersion. We perform full-wave numerical simulations to match the experimental results and use a semi-analytical approach based on the waveguide theory to provide clear physical insights behind the observations.
\begin{figure}[ht]
    \centering
\includegraphics[width=0.95\linewidth]{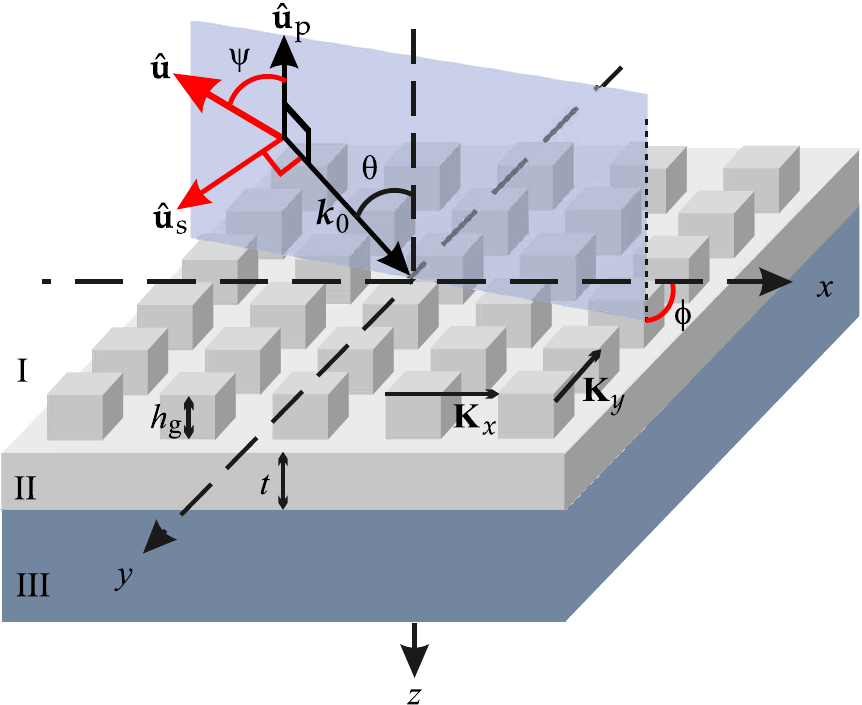}
    \caption{Schematic of a square-lattice GMRF with periodicities along $x$ and $y$ under conical illumination. The incident harmonic wave is linearly polarized with the electric field pointing along $\mathbf{\hat u}$.}
    \label{fig:1}
\end{figure}

The geometry of our GMRF is depicted in Fig.~\ref{fig:1}, which consists of a 2-D periodic surface-relief binary grating on top of a thin waveguiding layer. The incident and substrate regions are homogeneous media with refractive indices $n_{\mathrm i}$ (air) and $n_{\mathrm s}$ (fused silica), respectively. The grating and the waveguide are made of the same silicon nitride (SiN$_{\mathrm{x}}$) material with refractive index $n_{\mathrm g}$. Grating periodicities along $x$- and $y$-axes are $\Lambda_x$ and $\Lambda_y$, and the grating line widths are $f_x\Lambda_x$ and $f_y\Lambda_y$ respectively. A linearly polarized harmonic wave with wave number $k_0$ is incident from the air at an angle of incidence (AOI) $\theta$ with $z$-axis.
The POI highlighted in blue color makes a conical angle $\phi$ with the $x$-axis. Unit vector characterizing polarization is denoted by $\mathbf{\hat u}$, and $\psi$ is the polarization angle. For the conical incidence of any plane wave, the wave vector $\mathbf{k_0}$ can be written as
\begin{equation}
\begin{split}
 \label{eq1}
 \mathbf{k}_0&=k_{0x}\mathbf{\hat{x}}+k_{0y}\mathbf{\hat{y}}+k_{0z}\mathbf{\hat{z}}\\
 &=k_0(\sin\theta\cos\phi\mathbf{\hat{x}}+\sin\theta\sin\phi\mathbf{\hat{y}}+\cos\theta\mathbf{\hat z}).
 \end{split}
\end{equation}

\noindent
Upon incidence on the GMRF, the plane wave gets diffracted, and the wave vectors of the diffracted light can be expressed as
\begin{multline}
 \label{eq3}
\mathbf{k}_{q,mn}=k_{xm}\hat{\mathbf{x}}+k_{yn}\hat{\mathbf{y}}+k_{zq,mn}\hat{\mathbf{z}}\\=(k_{0x}+mK_x)\hat{\mathbf{x}}+(k_{0y}+nK_y)\hat{\mathbf{y}}+\sqrt{(k_q^2-k_{xm}^2-k_{yn}^2)}\hat{\mathbf{z}},
\end{multline}
where $q=1,3$ stands for regions I and III (the superstrate and the substrate as illustrated in Fig.~\ref{fig:1}), $m$ and $n$ are integers representing the diffraction orders along $x$ and $y$-axes, respectively. The grating vectors are $\mathbf{K}_x=(2\pi/\Lambda_x)\mathbf{\hat{x}}$, and $\mathbf{K}_y=(2\pi/\Lambda_y)\mathbf{\hat{y}}$.

\begin{figure}[ht]
    \centering
    \includegraphics[width=\linewidth]{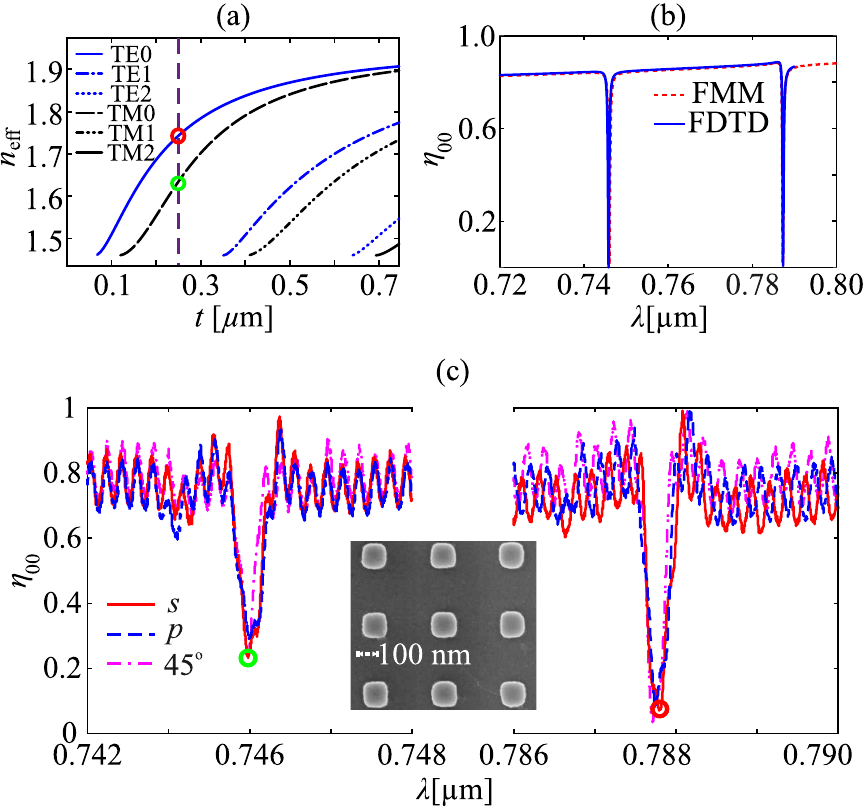}
    \caption{(a) Effective indices ($n_{\mathrm{eff}}$) of the modes versus waveguide thickness $t$ \cite{EIMS}. The wavelength (in the air) of the incident harmonic wave is $\lambda_0=0.744$ $\mu$m, $n_s=1.46$, and $n_g=1.95$ are used in the calculations. (b) diffraction efficiency in transmission calculated with the FMM and the FDTD-based solvers for $s$, $p$, or $45^{\circ}$ polarized harmonic wave in normal incidence from the air. $\Lambda_x$=$\Lambda_y$=0.451 $\mu$m, $f_x$=$f_y$=0.46, $h_{\mathrm g}$=0.045 $\mu$m, and $t$=0.245 $\mu$m are used in the numerical simulations, (c) Diffraction efficiency in direct transmission obtained experimentally for normal incidence of $s$, $p$, and $45^{\circ}$ polarized light. The inset shows a scanning electron microscope image (top-view) of the fabricated GMRF.}
    \label{fig:2}
\end{figure}

For a high-Q GMRF, the fields remain tightly confined inside the waveguide layer. Therefore, one can use the theory of an unperturbed slab waveguide to estimate the spectral peak positions of the resonances accurately. We use a multilayer waveguide mode solver \cite{EIMS} to evaluate the effective refractive indices $n_{\mathrm{eff}}$ of the modes in the SiN$_{\mathrm{x}}$ slab waveguide layer. Effective indices of the modes are plotted against the thickness ($t$) of the waveguide in Fig.~\ref{fig:2}(a). The plots show that the thickness of the waveguide layer limits the maximum number of supported modes. These modes can be classified either as TE, where  $\mathbf{E}$-field, or TM, where $\mathbf{H}$-field are perpendicular to the directions of propagation of the modes. We chose a waveguide layer thickness of $t=0.245$ $\mu$m (violet vertical dashed line in Fig.~\ref{fig:2}(a)) for the experimental demonstration. Clearly, this waveguide thickness can support one TE$_0$ (fundamental TE) and one TM$_0$ (fundamental TM) mode with effective mode indices $n^{\mathrm{TE}_0}_{\mathrm{eff}}$=1.7381, and $n^{\mathrm{TM}_0}_{\mathrm{eff}}$=1.6269, respectively.  
The propagation constant $\beta$ of the supported waveguide modes is defined as $\beta=n_{\mathrm{eff}}k_0$.

The resonance anomalies are observed when the grating diffraction matches the tangential wave number of the incident wave with the propagation constant of a slab waveguide mode. This condition. also known as the phase-matching condition, can, in general, be written as
\begin{equation}
\label{eq:4}
    \left|k_{xm}\hat{\mathbf x}+k_{yn}\hat{\mathbf y}\right|=\beta=n_{\mathrm{eff}}k_0.
\end{equation}
Using Eqs.~\eqref{eq1} and \eqref{eq3} it can be rewritten in the following form
\begin{equation}
\label{eq:5}
\left(k_0\sin\theta\cos\phi+\frac{2\pi m}{\Lambda_x}\right)^2+\left(k_0\sin\theta\sin\phi+\frac{2\pi n}{\Lambda_y}\right)^2=\beta^2.
\end{equation}

We first analyze the normal incidence scenario with the $xz$-plane as the POI ($\theta=0^{\circ}$ and $\phi=0^{\circ}$). For $s$-polarization, the $\mathbf{E}$-field is parallel to $y$-axis and projection of $\mathbf{k}_0$ on $xy$- plane is parallel with $x$-axis. For a better understanding, we can think of this complete diffraction problem as a combination of diffraction from two 1-D gratings (one grating with periodicity along the $x$-axis and another with periodicity along the $y$-axis). Hence, the grating periodic along the $x$-axis is under classical diffraction mounting for $s$-polarized light incidence. At the same time, the $y$-periodic grating is under full-conical illumination.
For normal incidence, Eq.~\eqref{eq:5} reduces to
\begin{equation}
\label{eq:7}
\sqrt{\left(\frac{2\pi m}{\Lambda_x}\right)^2+\left(\frac{2\pi n}{\Lambda_y}\right)^2}=\pm \frac{2\pi}{\lambda_0}n_{\mathrm{eff}}^{\mathrm{TE/TM}}.
\end{equation}
The diffraction orders $(m,n)=(\pm 1,0)$ excite the TE$_0$ waveguide modes propagating along $\pm x$-directions, and the spectral peak position of the resonance ($\Gamma$-point) is at $\lambda_{0}^{\mathrm{TE}_0}=0.788$ $\mu$m, which is obtained  by using $n^{\mathrm{TE}_0}_{\mathrm{eff}}$=1.7381 in Eq.~\eqref{eq:7}. Simultaneously, the $(0,\pm 1)$ diffraction orders excite the TM$_0$ waveguide modes at $\lambda_{0}^{\mathrm{TM}_0}=0.744 $ $\mu$m ($\mathrm F$-point)  propagating along $\pm y$-directions. 

To validate the preceding analytical predictions of the spectral positions of the resonances, we perform full-wave numerical simulations with in-house software based on the well-known Fourier Modal Method (FMM) \cite{Li:97}. The FMM-based simulation results are also compared with Finite-Difference Time-Domain (FDTD) simulations based on a commercial solver \cite{FDTD}. For both methods, careful convergence tests are performed to ensure the accuracy of the numerical results (see section 1 of the supplementary information). Furthermore, an experimentally measured refractive index data of SiN$_{\mathrm{x}}$ is used in the numerical (FMM and FDTD) simulations (see section 2 of the supplementary information). The transmittance spectrum for illumination with a $s$-polarized harmonic wave is shown in Fig.~\ref{fig:2}(b). For normal incidence, the spectrum is polarization-independent. Hence, one can obtain identical simulation results with $p$ and $45^{\circ}$ linear polarizations.

For experimental demonstration, a square-lattice GMRF with the designed grating parameters is fabricated with standard lithographic processes, which include steps such as chemical vapor deposition, electron beam lithography, and reactive ion etching (see section 3 of the supplementary information for the fabrication details). The transmission spectra of the GMRF are measured with a custom-built transmission setup. A supercontinuum light source is used, and the transmittance in $0$-th diffraction order is measured with an optical spectrum analyzer (OSA). The GMRF is mounted on a 2-axis goniometer placed on top of a rotation stage to control both $\theta$ and $\phi$ (see section S4 in the supplementary information for details of the experimental procedure). The normalized efficiency in direct transmission ($\eta_{00}$) for $s$, $p$, and $45^{\circ}$ linear polarizations of light incident from the air with $\theta=0^{\circ}$, and $\phi=0^{\circ}$ are shown in Fig.~\ref{fig:2}(c). The Q-factors of the resonances estimated from the experimental plots in Fig.~\ref{fig:2}(c) are $\approx 4500$. Figures~\ref{fig:2}(a)-(c) show that the analytical predictions, full-wave numerical simulations, and experimental results are in excellent agreement. 

\begin{figure}[ht]
    \centering
    \includegraphics[width=\linewidth]{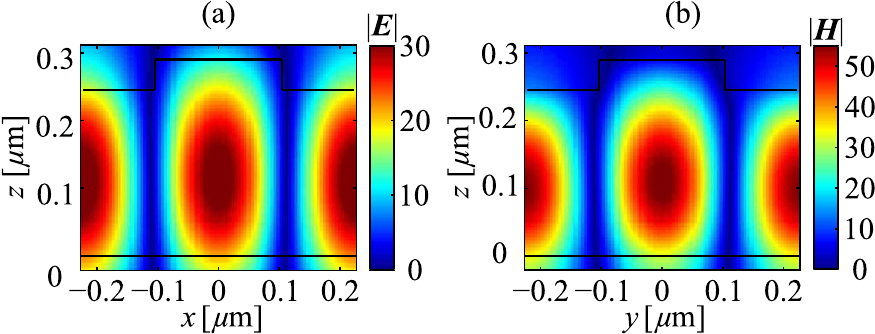}
    \caption{Fields inside the GMRF at resonance. (a) the $\mathbf{E}$ field inside the GMRF normalized to the incident EM wave for $\lambda_{\Gamma}$, and (b) the normalized $\mathbf{H}$ field inside the grating at $\lambda_{{\mathrm F}}$.}
    \label{fig:3}
\end{figure}

We calculate the ﬁeld distributions inside the GMRF with the FDTD-based solver to confirm the analytical predictions of the mode propagation directions. Figure~\ref{fig:3} shows the simulation results. The excitation wavelength is set as $\Gamma_{s}=0.7878$ $\mu$m for the $E$-field plot in Fig.~\ref{fig:3}(a). For the $H$-field plot in Fig.~\ref{fig:3}(b), the excitation wavelength is ${\mathrm F}_s=0.7457$ $\mu$m. In both plots, $s$-polarized light is incident normally onto the GMRF from the air. The plots show that excitations of the guided modes result in strong confinements and enhancements of the ﬁelds within the waveguide. Moreover, the TE$_0$ and the TM$_0$ modes propagate along the $x$ and the $y$-axes, respectively.
\begin{figure*}[ht]
    \centering
    \includegraphics[width=\linewidth]{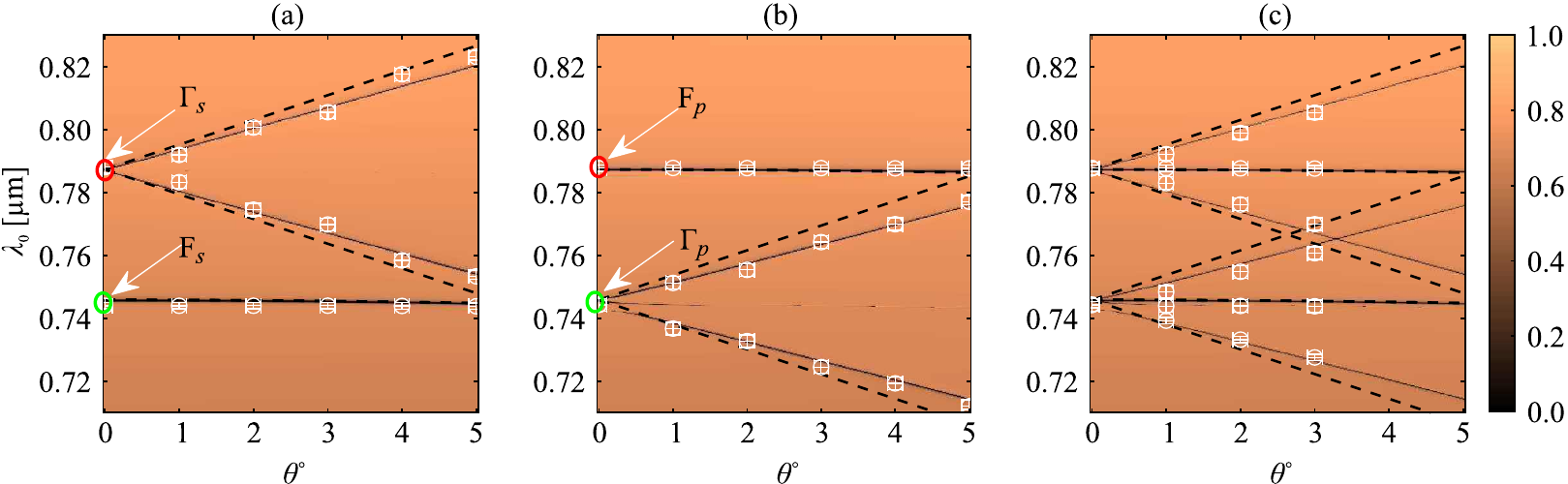}
    \caption{Diffraction efficiency in direct transmission plotted against $\lambda_0$, and $\theta$ for (a) $s$-polarized ($\psi=\pi⁄2$), (b) $p$-polarized ($\psi=0$), and (c) $45^{\circ}$ linearly polarized illumination ($\psi=\pi⁄4$), respectively. The continuous lines indicate the FMM solver-based numerical simulation results, the dispersion plots obtained with Eq.~\eqref{eq:5} are indicated with the dashed lines, and the white circles with error bars mark the experimental results. The color bar scaled from '0' to '1' indicates the transmittance value.}
    \label{fig:4}
\end{figure*}

We proceed to investigate the angular and spectral dispersion characteristics of the resonances. For the oblique incidence of a harmonic wave ($\theta\neq 0^{\circ}$) and for $\phi=0^{\circ}$, Eq.~\eqref{eq:5} can either take the form of Eq.~\eqref{eq:5a} or Eq.~\eqref{eq:5b}, which solely depends on the indices ($m,n$) of the diffraction orders that excite the guided modes. For $s$-polarized light, the $(m,n)=(\pm 1,0)$ diffraction orders of the grating excite the TE$_0$ waveguide modes propagating along $\pm x$-directions. Hence, the resonances related to the TE$_0$ modes symmetrically split into two branches, as shown by Eq.~\eqref{eq:5a}.
\begin{equation}
\label{eq:5a}
\pm k_0\sin\theta+\frac{2\pi m}{\Lambda_x}=\pm \frac{2\pi}{\lambda_0}n_{\mathrm{eff}}^{\mathrm{TE}}.
\end{equation}
The resulting V-shaped dispersion is plotted in Fig.~\ref{fig:4}(a) as the dashed lines extending from $\Gamma_{s}=0.788$ $\mu$m. Simultaneously, the $(m,n)=(0,\pm 1)$ diffraction orders can excite both TM$_0$ and TE$_0$ waveguide modes. For these degenerate modes, Eq.~\eqref{eq:5} can be written as 
\begin{equation}
\label{eq:5b}
\sqrt{(k_0\sin\theta)^2+\left(\frac{2\pi n}{\Lambda_y}\right)^2}=\pm\frac{2\pi}{\lambda_0}n_{\mathrm{eff}}^{\mathrm{TM/TE}},
\end{equation}
which shows no splitting of the nearly flat dispersion curve as illustrated with the dashed line originating from $\mathrm F_{\mathrm{s}}=0.744$ $\mu$m (for the TM mode) in Fig.~\ref{fig:4}(a). It should be noted that the near-flat TE modes excited by the $(m,n)=(0,\pm 1)$ diffraction orders propagate along $k_x\hat{x}\pm K_y\hat{y}$. For small values of $\theta$, the coupling of $s$-polarized light with these TE-modes is small due to the polarization symmetry (see section 5 in the supplementary information). At normal incidence, this coupling possibility vanishes completely, resulting in band gaps.
 
 For $p$ polarized light, the $(m,n)=(\pm 1,0)$ diffraction orders of the grating excite the TM$_0$ guided modes resulting in the V-shaped dispersion curve starting from $\Gamma_{p}$. Simultaneously, both TM$_0$ and TE$_0$ modes are excited with the $(m,n)=(0,\pm 1)$ diffraction orders displaying nearly flat dispersion characteristics as shown in Fig.~\ref{fig:4}(b).
 
 Any arbitrary linear polarization state of light can be considered a sum of the $s$ and $p$ polarized components. For values of $\psi$ other than $0^{\circ}$ or $90^{\circ}$, both the $(m,n)=(\pm 1, 0)$, and $(m,n)=(0,\pm 1)$  diffraction orders of the grating can excite TE$_0$/TM$_0$-type slab waveguide modes efficiently. The dashed lines in Figure~\ref{fig:4}(c) show the angular/ spectral dispersion of the modes for the oblique incidence of a 45$^\circ$ polarized harmonic wave. Four branches of the dispersion curves can be observed for $\theta\neq 0^{\circ}$. Two are V-shaped, whereas the other two are almost horizontal lines. Hence, the dispersion characteristics for 45$^\circ$ linear polarization case can be seen as a combination of the $s$ and $p$ polarized cases. 
 
 The numerically calculated transmission spectra with FMM are also included in Fig.~\ref{fig:4}. The color bar indicates the normalized transmittance ($\eta_{00}$) in direct transmission. We notice that the $\Gamma$ and the $\mathrm{F}$ points obtained with the two approaches agree very well. However, as $\theta$ increases, the dispersion curves obtained with the analytical formulation start to deviate from those obtained with FMM. This difference can be attributed to wavelength-dependent mode indices. The analytical solution of the dispersion curve in Eq.~\eqref{eq:5} assumes a constant $n_{\mathrm{eff}}$. However, a modified $n_{\mathrm{eff}}(\lambda_0)$ should be considered for a more accurate analytic estimation.

 The experimentally measured peak wavelengths of the transmittance curves at discrete $\theta$ are also included in Fig.~\ref{fig:4}. The red (green) and green (red) circle positions correspond to $\Gamma$ and $\mathrm{F}$ points for $s$ ($p$) polarization. The experimentally obtained resonance peak positions $\Gamma_{s}=0.787$ $\mu$m, and $\mathrm{F}_{s}=0.744$ are in excellent agreement with the simulation results. The error bars along $\lambda_0$ and $\theta$ axes are related to the measurement uncertainties (see section 6 of the supplementary information).

In summary, we have identified and experimentally verified a type of high-Q guided-mode resonance with ultra-low spectral and angular dispersion in a 2-D grating. The dispersion index ($\Delta\lambda_0/\Delta\theta$) of this resonance with Q-factor $\approx4500$ is $\sim10^{-5}$ $\mu$m/degree, which is about three orders of magnitude smaller than the dispersion of the other resonant mode (with a similar Q-factor) that can be excited in the 2-D GMRF simultaneously. Besides, we experimentally demonstrate that one can swap the dispersion characteristics of these two resonances by switching from $s$ to $p$ polarized light incidence and vice versa. The results presented in this letter show that achieving high-Q and low dispersion simultaneously with a free-space diffractive optical element is possible and will benefit many practical applications in optoelectronics and photonics that rely on resonances from free-space-coupled geometries.

\begin{acknowledgments}
The authors thank Timo Stolt and Alireza Rahimi Rashed for the helpful discussions during the planning stage of the optical measurements.
\end{acknowledgments}

\section*{Author contribution}
SB and TN conceived the idea and planned the research. RA, SB, and HP performed the numerical simulations. SB and PK fabricated the structure. RA and SB performed the optical measurements. RA analyzed the experimental data. SB, TN, and RA prepared the manuscript. SB, TN, and JT supervised the project. All authors edited the final version of the manuscript.

\section*{Funding}
This work is supported by the Flagship of Photonics Research and Innovation (PREIN) funded by the Academy of Finland- grant nos. 31001498, 320165 (Tampere University ), and 320166 (University of Eastern Finland).

\section*{Data Availability Statement}
The data that support the findings of this study are available from the corresponding author upon reasonable request.
\section*{Supplementary information}
See the Supplementary file for supporting contents.

\nocite{*}
\bibliography{aipsamp}

\end{document}